\documentclass[twocolumn,aps,superscriptaddress]{revtex4}
\usepackage{latexsym,amssymb,amsmath}
\usepackage{color}
\usepackage[bookmarksnumbered,bookmarksopen,colorlinks,citecolor=red,linkcolor=blue]{hyperref}
\usepackage{mathrsfs}
\usepackage{times}
\usepackage{csquotes}
\usepackage{amssymb}
\usepackage{amsmath}
\usepackage{graphicx}
\usepackage{subfig}
\usepackage[normalem]{ulem}

\begin{document}

\title{Finding signatures of the nuclear symmetry energy in heavy-ion collisions with deep learning}

\author{Yongjia Wang\footnote{Corresponding author: wangyongjia@zjhu.edu.cn}}
\affiliation{School of Science, Huzhou University, Huzhou 313000, China}
\author{Fupeng Li}
\affiliation{School of Science, Huzhou University, Huzhou 313000, China}
\affiliation{School of Science, Zhejiang University of Technology, Hangzhou 310014, China}
\author{Qingfeng Li\footnote{Corresponding author: liqf@zjhu.edu.cn}}
\affiliation{School of Science, Huzhou University, Huzhou 313000,  China}
\affiliation{Institute of Modern Physics, Chinese Academy of Sciences, Lanzhou 730000, China}

\author{Hongliang L\"{u}}
\affiliation{HiSilicon Research Department, Huawei Technologies Co., Ltd., Shenzhen 518000, China}
\author{Kai Zhou}
\affiliation{Frankfurt Institute for Advanced Studies, Ruth-Moufang-Str. 1, 60438 Frankfurt am Main, Germany}

\begin{abstract}
A deep convolutional neural network (CNN) is developed to study symmetry energy ($E_{\rm sym}(\rho)$) effects by learning the mapping between the symmetry energy and the two-dimensional (transverse momentum and rapidity) distributions of protons and neutrons in heavy-ion collisions.
Supervised training is performed with labelled data-set from the ultrarelativistic quantum molecular dynamics (UrQMD) model simulation. It is found that, by using proton spectra on event-by-event basis as input, the accuracy for classifying the soft and stiff $E_{\rm sym}(\rho)$ is about 60\% due to large event-by-event fluctuations, while by setting event-summed proton spectra as input, the classification accuracy increases to 98\%. The accuracy for 5-label (5 different $E_{\rm sym}(\rho)$) classification task are about 58\% and 72\% by using proton and neutron spectra, respectively. For the regression task, the mean absolute error (MAE) which measures the average magnitude of the absolute differences between the predicted and actual $L$ (the slope parameter of $E_{\rm sym}(\rho)$) are about 20.4 and 14.8 MeV by using proton and neutron spectra, respectively. Fingerprints of the density-dependent nuclear symmetry energy on the transverse momentum and rapidity distributions of protons and neutrons can be identified by convolutional neural network algorithm.

\end{abstract}

\pacs{21.65.Ef, 21.65.Mn, 25.70.-z}

\maketitle
\section{Introduction}

The density-dependent nuclear symmetry energy $E_{\rm sym}(\rho)$ has attracted considerable attention from both nuclear physics and nuclear astrophysics communities in the recent two decades, since its knowledge is crucial for our understanding of diverse phenomena observed in rare isotopes, nuclear reactions with exotic nuclei, as well as neutron star and its merger~\cite{BALi08,Tsang:2012se,Baldo:2016jhp,Oertel:2016bki,Li:2018lpy,Roca-Maza:2018ujj,Gao:2019vby}. Constraints on $E_{\rm sym}(\rho)$ with various observations from studies of nuclear structure, nuclear reaction, and neutron star properties have been studies. However, the whole picture of nuclear symmetry energy as a function of density is still indistinct, especially above the saturation density ($\rho_0$).
Simulations with transport model in combination with observables in heavy-ion collisions is one of the important way to constrain the high density behavior of $E_{\rm sym}(\rho)$. Several sensitive observables have been presented, but it is still difficult to get a tight and consistent constraint on $E_{\rm sym}(\rho)$ at high densities ~\cite{Colonna:2020euy,Ono:2019jxm,Xu:2019hqg,FOP-wyj,FOP-zyx,Ma:2018wtw,Feng:2018emx}.

Deep learning has been proved useful for analyzing pattern from complex data in many branches of science, such as in physics~\cite{ROMP,XLF,HL,Nature1,Steinheimer:2019iso,Song:2021rmm,Wang:2020tgb,Ma:2020bic,Ma:2020mbd,PJC,Utama:2016tcl,Zhou:2018ill,Shi:2021qri}.
In heavy-ion physics, neural network has been used to determine the impact parameter in heavy-ion collisions since 1990s ~\cite{CDPRC,SBASS,Haddad,JD,Li:2020qqn,Kuttan:2020kha,lifup2}. The convolutional neural network (CNN) which is successful in Computer Vision has been shown promising in studying Quantum chromodynamics (QCD) properties from heavy-ion observables ~\cite{PLG,Du:2019civ,Kvasiuk:2020izb,Thaprasop:2020mzp,Zhao:2021yjo}. The marriage of heavy-ion physics and deep learning brought new effective paradigm for studying various details of the underlying physics.
In this work we attempt to find fingerprints of the density-dependent nuclear symmetry energy in heavy-ion collisions by using a deep learning algorithm. This is a challenge task because symmetry energy is a sub-leading ingredient of transport model for studying heavy-ion collisions at intermediate energies. In addition, symmetry energy effects may be further washed out by the stochastic nucleon-nucleon interactions.

\section{Model and data description}

\label{model}

\begin{table}[htbp]
\caption{\label{tab:table1} The nuclear symmetry
energy coefficient $S_0$, the slope $L$, and the curvature $K_{sym}$ as obtained with
the five Skyrme interactions used in this work.}
\setlength{\tabcolsep}{1.2pt}
\begin{ruledtabular}
\begin{tabular}{lcccccccc}
&$S_0$(MeV)
&$L $(MeV)
&$K_{sym}$(MeV) \\
\hline
Skz4	&	32.0	&	5.8	&	-240.9	\\
SLy230a	&	32.0	&	44.3	&	-98.2	\\
SV-sym34 &	34.1	&	81.2	&	-79.7	\\
SkI2	&	33.9	&	106.4	&	73.2	\\
SkI1	&	37.2	&	159.0	&	229.1	\\
\end{tabular}

\end{ruledtabular}\label{skyrme}
\end{table}

\begin{figure}[htbp]
\centering
\includegraphics[angle=0,width=0.4\textwidth]{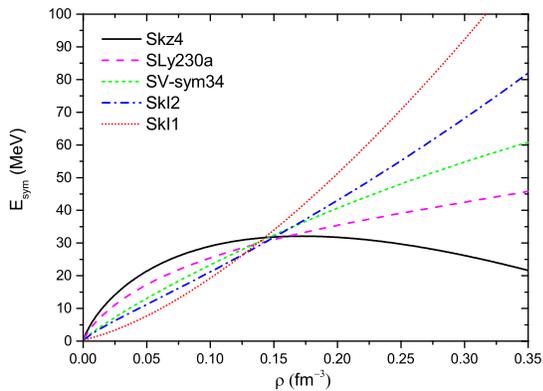}
\caption{\label{fig1}(Color online) The nuclear symmetry energy $E_{sym}(\rho)$ as a function of density. The various lines show predictions for the selected 5 Skyrme interactions in Tab. \ref{skyrme}.}
\end{figure}

It is known that deep learning algorithm relies on data heavily for pattern recognition. For our purpose, the data can be either from experiment or theoretical simulations. As data from theoretical simulations can be well controlled to perform supervised learning, in this work we apply the ultrarelativistic quantum molecular dynamics (UrQMD) model to generate training data. As a many-body microscopic transport model, UrQMD has been widely employed for investigating HIC from the Fermi energy (40 MeV per nucleon) up to the CERN Large Hadron Collider energies (TeV). With further improvement on several ingredients in UrQMD simulation, such as, the nuclear mean-field potential, the collision term, and the cluster recognition term, many experimental data within a wide energy range can be reproduced \cite{SAB,BLE,qfli1,qfli2,FOP-wyj,FOP-zyx}. In presently used UrQMD model, the symmetry potential is derived from the Skyrme potential energy density functional in the same manner as the improved quantum molecular dynamics (ImQMD) model, see e.g., Refs.~\cite{Zhang:2006vb,Wang:2013wca}. While the isoscalar components of the mean field potential inherits the widely used soft and momentum dependent version of potential in QMD-like models \cite{Aichelin:1991xy,Hartnack:1997ez}. As the high-density behavior of nuclear symmetry energy is still not well constrained, five different Skyrme interactions which yield very different $E_{\rm sym}(\rho)$ are considered in the present work, as plotted in Fig. \ref{fig1} and listed in Tab. \ref{tab:table1}. The slope of the density-dependent nuclear symmetry energy which is defined as $L=3\rho_{0}\left(\frac{\partial{E_{\rm sym}(\rho)}}{\partial\rho}\right)|_{\rho=\rho_{0}}$) spans the range from 5 MeV up to 160 MeV, and covers a wide range of various constraints on $L$. As the overall contributions of the isovector part in HICs is relatively small compared to the
isoscalar part of the nuclear potential, the subtle influence on various observables hard to be revealed. Usually, the difference or ratio of observables between isospin partners may provide some hints for the isovector part of the nuclear potential.


\begin{figure}[t]
\centering
\includegraphics[angle=0,width=0.48\textwidth]{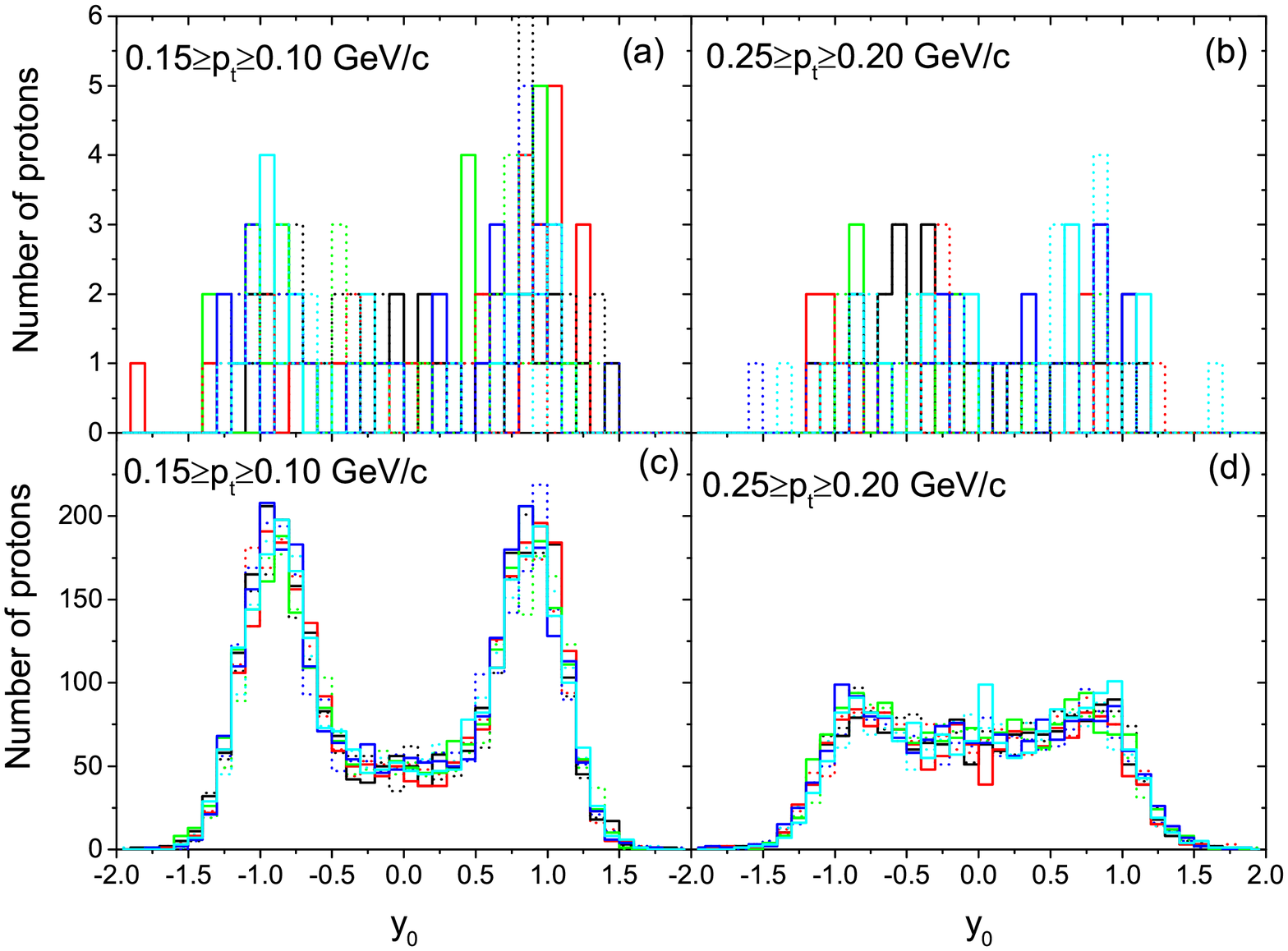}
\caption{\label{fig4}(Color online) Proton rapidity distribution calculated with Skz4 (solid lines) and SkI1 (dotted lines) interactions. In upper panels, results from five random events are displayed for each interactions. In lower panels, results from five random samples (which obtained by combining 100 collision events) for each interactions are displayed. }
\end{figure}

For data generation, 800 000 Au+Au collision events with impact parameter $b$=5 fm and beam energy $E_{\rm lab}$=0.4 GeV$/$nucleon for each symmetry energy are simulated, with the transverse momentum $p_t$ and rapidity $y_0$ of protons and neutrons all recorded. Due to initial fluctuations and the random nucleon-nucleon collisions, fluctuations on the rapidity and transverse momentum distributions are very large, consequently, the effects of symmetry energy on the distributions are hidden. As shown in Fig.\ref{fig4} where the proton rapidity distribution in $0.10 \leq p_t \leq 0.15$ GeV$/$c and $0.20 \leq p_t \leq 0.25$ GeV$/$c obtained from calculations with very soft (SKz4) and stiff (SkI1) symmetry energies are compared. Fluctuation on the distributions can be reduced by combining results from different events, as displayed in the lower panel of Fig.\ref{fig4}, nevertheless, the differences in the proton distributions obtained from these two symmetry energies are still too faint to be distinguished by conventional analysis. Accordingly, we prepare input sample by combining proton spectra from 100 UrQMD events. Therefore, we have 8 000 samples for each symmetry energy to perform supervised learning. We note here that, the training accuracy will increase if we combine more events into an input sample because of the fluctuation reduction. However, given a certain number of events, combining more events to input will reduce the number of training samples, which may depress the performance of CNN training. Combining 100 events to a sample is a compromise between fluctuation and the number of training samples.

The CNN architecture used in this work is inspired from successful applications in Refs\cite{PLG,Li:2020qqn,Du:2019civ}, which has two convolution layers and one subsequent fully-connected layer. The input to the CNN is the two-dimensional (transverse momentum and rapidity) distributions of protons, which is a 20$\times$40 matrix, as the transverse momentum $p_t$ spans from 0 to 1 GeV/c with 20 bins and the rapidity $y_0$ spans from -2 to 2 with 40 bins in between, respectively. A batch normalization (only after the first layer), LeakyReLU activation with a slope of 0.1, dropout with a rate of 0.2, as well as average pooling with a kernel size 2$\times$2 and a stride of 2 pixels are added between the two layers. Each convolution layer consist of 128 filters of size 5$\times$5. We have checked that, the accuracy is hardly influenced when varying the above mentioned parameters or adding more layers, these variations only affect the runtime and the stability of training process.

\section{results and discussion}
\subsection{Result of two-class classification task}

\begin{figure}[htbp]
\centering
\includegraphics[angle=0,width=0.48\textwidth]{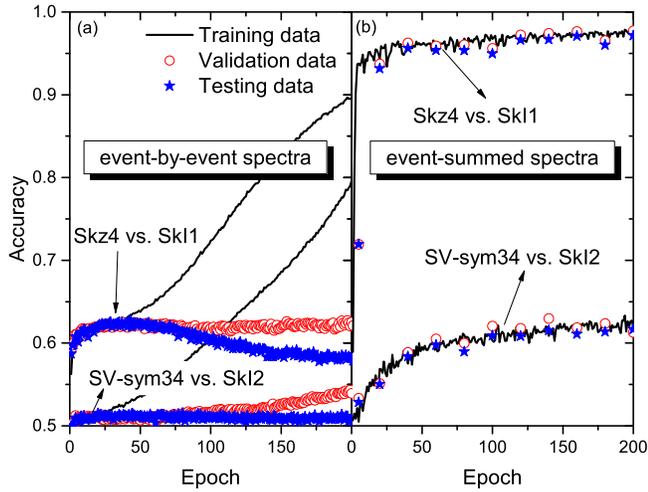}
\caption{\label{fig5}(Color online) The accuracy of two-class classification task as a function of the epoch (training time). Left panel: the accuracies for classifying Skz4 and SkI1, SV-sym34 and SkI2 by using classifier trained with event-by-event proton spectra. Right panel: similar as the left plot but with event-summed spectra as input for training and testing. In each plot, the accuracies for training, validation and testing data are displayed.}
\end{figure}

\begin{figure}[htbp]
\centering
\includegraphics[angle=0,width=0.4\textwidth]{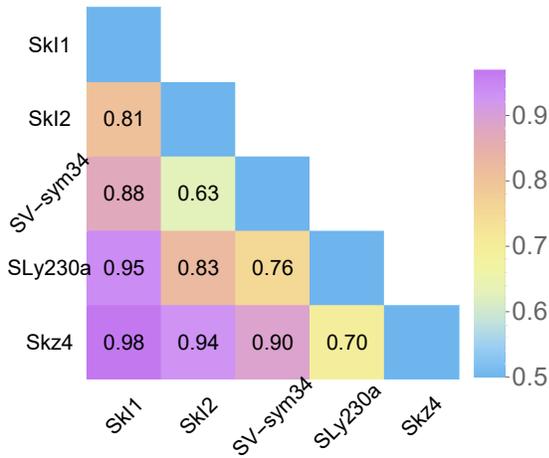}
\caption{\label{fig6}(Color online) The accuracy of two-class classification task. The number in each cell denotes the accuracy for classifying the vertical and horizontal labelled symmetry energies. The statistical error of the accuracy was estimated to be smaller than 2\% by comparing parallel testing data, being therefore negligible.}
\end{figure}

The simulated events for each symmetry energy are randomly divided into three parts: training, validation, and testing sets with the ratio of 60:15:25. Training set is used to adjust the parameters in CNN, validation set is used to monitor and avoid overfitting during the training by ensuring that the performance over both the training and validation set should not deviate a lot, otherwise the CNN is overfitting and the training should stop. Testing set is used to evaluate the actual predictive power of CNN on unseen different events. As CNN has a deep structure and consists of many parameters, in principle the training accuracy can keep on increasing to 100\%, while the validation and testing accuracy may not always increase unless the CNN learns the underlying relevant rules. As displayed in the left panel of Fig.\ref{fig5}, after 50 epochs the training accuracy still increase while the validation and test accuracy does not, indicating overfitting happening and accordingly the training should cease before epoch=50. The testing accuracy on distinguishing Skz4 and SkI1 is about 0.6 by using event-by-event proton spectra. For SV-sym34 and SkI2, the testing accuracy is about 0.5, meaning that the proton spectra obtained from these two symmetry energies are indistinguishable by the machine. The results by using event-summed samples (100 event-summed proton spectra) are displayed in that both testing and validation accuracy is enhanced. The accuracy is about 98\% and 63\% for classifying Skz4 vs SkI1, and SV-sym34 vs SkI2, respectively. As one expects, the former has larger accuracy because of the larger difference in $E_{\rm sym}(\rho)$ as shown in Fig.\ref{skyrme}. By averaging over 100 events, the fluctuation is reduced largely and the tiny difference on samples obtained from different symmetry energy can be partly identified by the machine. In HICs at intermediate energies, there are basically two sources of fluctuations: the initial fluctuation and dynamical fluctuation (i.e., stochastic particle collision). We have tried to reduce the initial fluctuation by artificially using the same initial nuclei in every collision events. Consequently, the accuracy for classifying Skz4 and SkI1 on the event-by-event basis reaches the range 70\%-90\%, which depends strongly on the random number generator seed. Fig.\ref{fig6} displays the accuracy for classifying two different symmetry energies by using event-summed proton spectra. The larger the difference in $E_{\rm sym}(\rho)$, the higher the accuracy for their classification. The accuracy for classifying SV-sym34 vs SkI2 is the lowest of all, since their difference in $L$ is the smallest.
Generally, the accuracy increases with the slope $L$ difference between two symmetry energies increasing, showing that the CNN is capable of manifesting fingerprints of the density-dependent nuclear symmetry energy on proton spectra.
\subsection{Result of five-class classification task}

\begin{figure}[htbp]
\centering
\includegraphics[angle=0,width=0.4\textwidth]{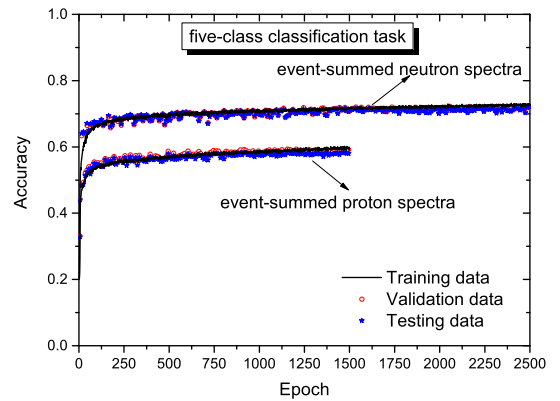}
\caption{\label{fig7}(Color online) The accuracy of five-class classification task as a function of the epoch with event-summed proton or neutron spectra used.}
\end{figure}

\begin{figure}[htbp]
\centering
\includegraphics[angle=0,width=0.4\textwidth]{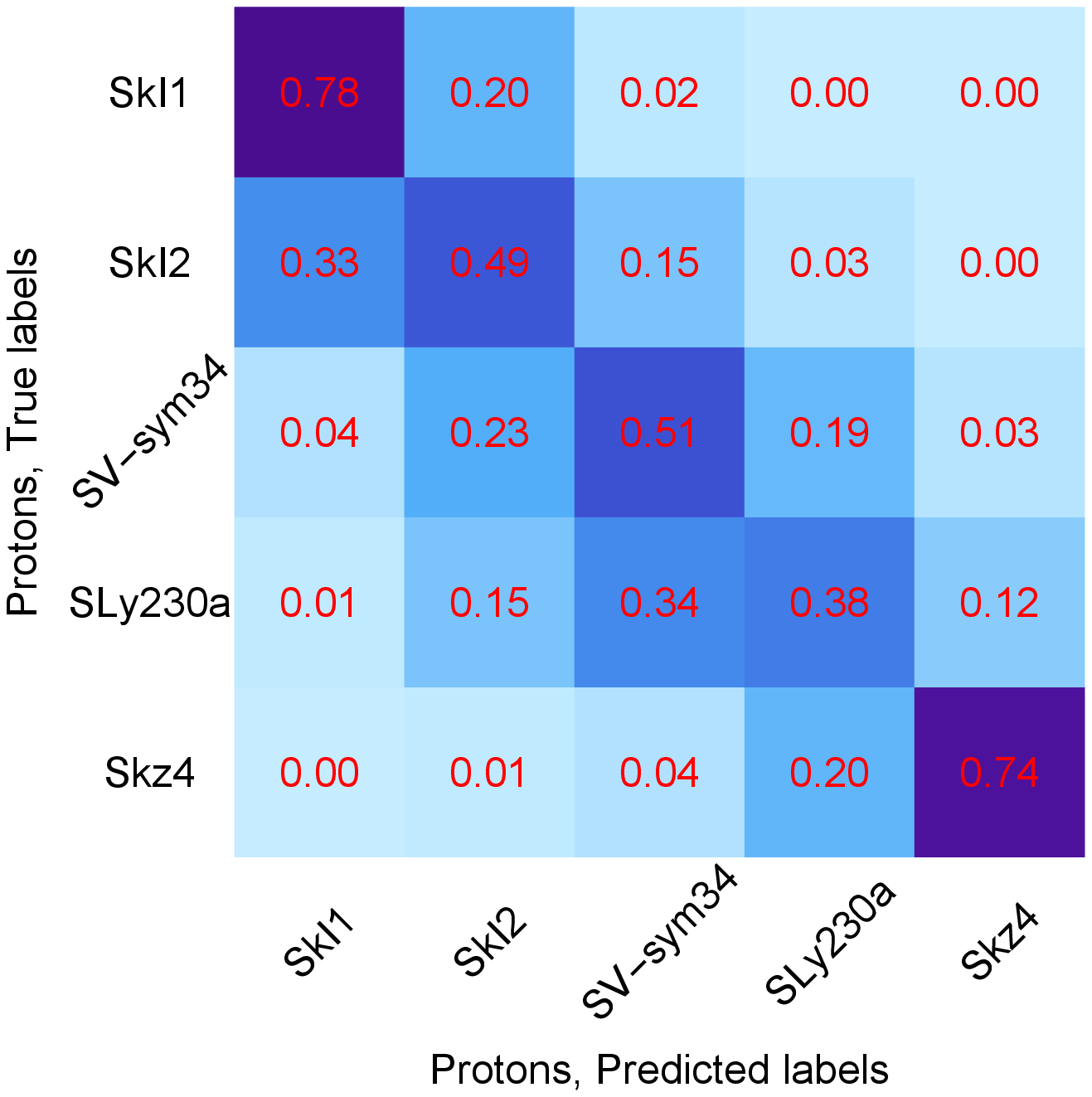}
\includegraphics[angle=0,width=0.4\textwidth]{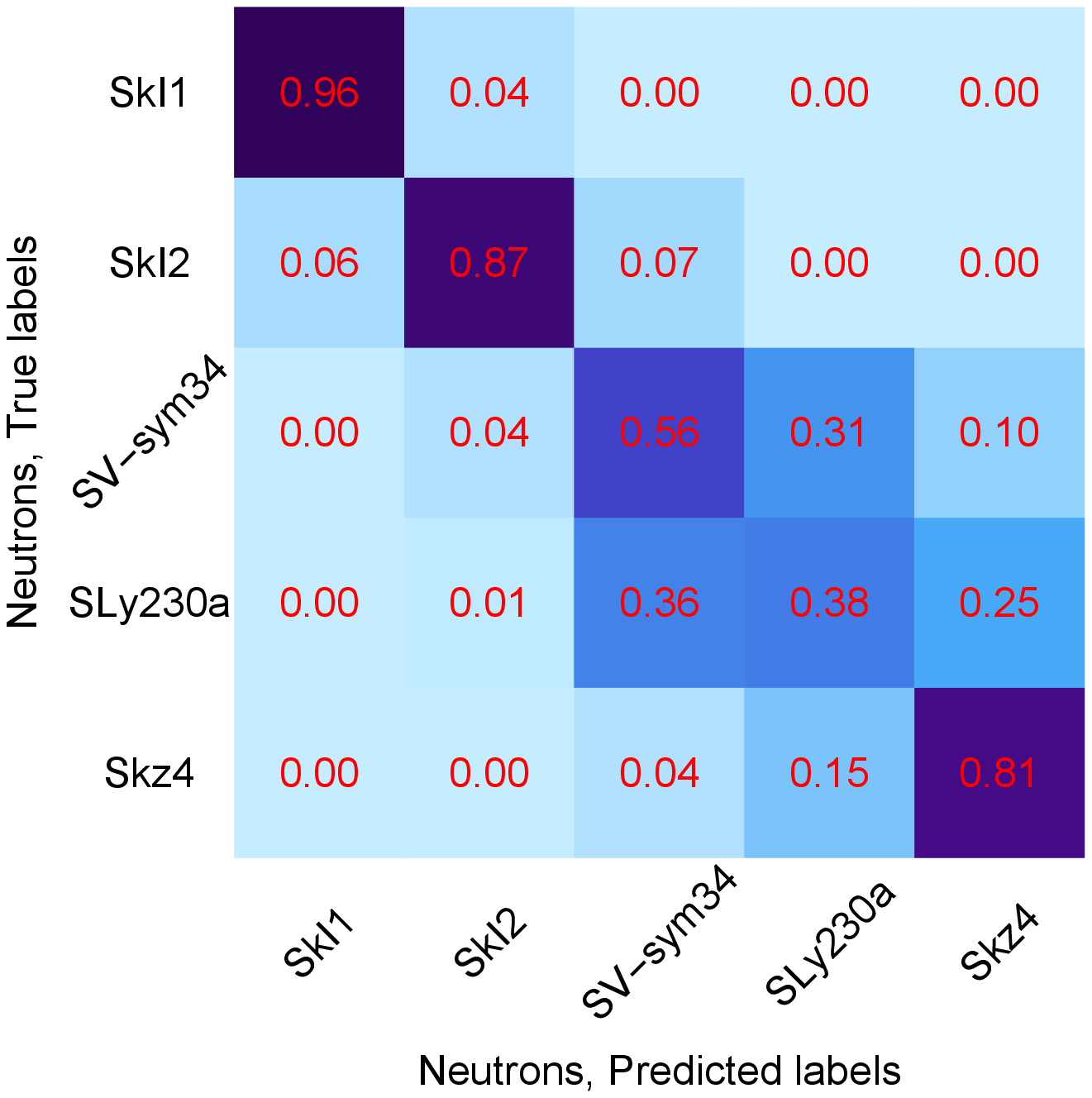}
\caption{\label{fig8}(Color online) The confusion matrix for five-class classification task. Number in each off-diagonal cell represents the probability that the object of the symmetry energy (vertical label) being misclassified as the horizontal labelled symmetry energy. The diagonal entries show the fraction of correctly classified testing data. Thus the sum of number in each row is unit. Upper and lower panels denote the results by using event-summed proton and neutron spectra, respectively.}
\end{figure}

The accuracy of the five-class classification task is plotted in Fig. \ref{fig7}, where the results by using event-summed either proton or neutron spectra are displayed. Using proton spectra, the accuracy is about 58\% which is almost three times that of a random guessing. While the accuracy is increased to 72\% if the neutron spectra is used. It is understandable that the accuracy with neutron spectra is higher because neutron-related observables are usually more sensitive to $E_{\rm sym}(\rho)$ than proton-related observables, see, e.g., Refs.\cite{Wang:2020dru,Wang:2014rva}.

The confusion matrix is a good way to display the performance of multi-class classification model in making prediction.
Fig. \ref{fig8} shows the confusion matrix for the five-class classification task. The diagonal numbers denote the probability that a horizontal labelled symmetry energy is correctly classified, as can be seen that they are the largest one in each row, although the probability is very high for some symmetry energies (e.g., SkI1 and Skz4) and very low for others (e.g., SV-sym34 and SLy230a). For example, by using event-summed proton spectra, 78\% of SkI1 sample can be correctly identified, while 20\% of them are misidentified as SkI2, and the remaining 2\% are misidentified as SV-sym34. This result is reasonable as the symmetry energies obtained with SkI1 and SkI2 are close to each other, thus the probability of misrecognizing each other is high. Indeed, numbers around diagonal are larger than others, meaning that the predicted labels are close to the ground truth, although the classifier cannot always give the correct answer, indicating that the CNN can indeed capture symmetry energy signals in the spectra.

\subsection{Result of regression task}

\begin{figure}[htbp]
\begin{centering}
\includegraphics[angle=0,width=0.48\textwidth]{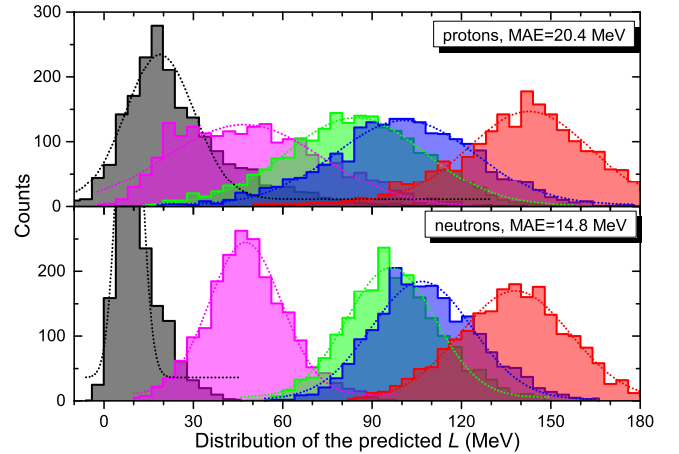}
\caption{\label{fig9}(Color online) The distribution of the predicted slope parameter $L$. 2000 samples are tested for each symmetry energy. Upper and lower panels denote the results by using event-summed proton and neutron spectra, respectively. Dotted lines are Gaussian fits to the prediction.}
\end{centering}
\end{figure}

\begin{table}[!htb]
\caption{The mean values of predicted $L$ and its standard deviation $\sigma$ obtained with Gaussian fit, in units of MeV. }
\label{lll}
\begin{tabular*}{7cm}{@{\extracolsep{\fill}}cc|cc|ccc}
\toprule
&  & \multicolumn{2}{|c|}{Proton spectra}  & \multicolumn{2}{c}{Neutron spectra}                    \\
\cline{2-6}
         & $L_{true}$ & $L_{pred}$  & $\sigma$  & $L_{pred}$  & $\sigma$  \\
\hline
Skz4     & 5.8              & 18.9             & 11.5           & 8.5              & 3.7            \\
SLy230a  & 44.3             & 46.9             & 26.5           & 47.5             & 12.1           \\
SV-sym36 & 81.2             & 84.5             & 23.3           & 96.4             & 14.8           \\
Skl2     & 106.4            & 100.9            & 23.6           & 106.7            & 17.0           \\
Skl1     & 159.0            & 142.6            & 20.7           & 138.1            & 18.5           \\
\toprule
\end{tabular*}
\end{table}

It is known that the slope $L$ is one of the important quantities that characterizes the behavior of density-dependent symmetry energy. The CNN architecture is also adapted to predict the slope $L$ (regression task) instead of classifying symmetry energies. The distribution of the predicted slope parameter $L$ with event-summed proton and neutron spectra are plotted in Fig.\ref{fig9}. Mean absolute error (MAE), which is the absolute difference between the true and the predicted values, is about 20.4 and 14.8 MeV for using event-summed proton and neutron spectra, respectively. As can be seen, the predicted $L$ distributions for Skz4 and SkI1 are well separated with each other, while the distributions for SV-sym34 and SkI2 are overlapping each other largely, which is due to the fact that $L$ difference in the former case is about 153 MeV but it is only about 25 MeV in the later case. Using Gaussian fit, one can get the mean value of the predicted $L$ and its standard deviation, which are listed in Tab. \ref{lll}. The mean values of predicted $L$ are close to the true values used in corresponding events, indicating again the strong capability of CNN in revealing fingerprints of $E_{sym}(\rho)$ on the transverse momentum and rapidity distributions of protons and neutrons.

\section{summary}
To summarize, we have presented the first attempt to find fingerprints of the nuclear symmetry energy in heavy-ion collision with deep leaning. The two-dimensional (transverse momentum and rapidity) distributions of protons and neutrons simulated with UrQMD model are fed into a CNN, and the output of CNN is either the label which denotes the stiffness of $E_{\rm sym}(\rho)$ or the slope parameter $L$. It is found that, when using proton distributions on event-by-event basis, the accuracy for classifying the soft and stiff $E_{\rm sym}(\rho)$ is about 60\%, due to large event-by-event fluctuations, while by using event-summed proton spectra as the input sample the accuracy increases to 98\%. For classifying five different $E_{\rm sym}(\rho)$, the accuracy is about 58\% and 72\% when proton and neutron spectra is used, respectively. For the $L$ regression, the mean absolute error between the CNN predicted and true $L$ are about 20.4 and 14.8 MeV by using proton and neutron spectra, respectively. The present results suggest fingerprints of $E_{sym}(\rho)$ on the transverse momentum and rapidity distributions can be identified by deep learning algorithm.

\emph{Acknowledgement}.-- We acknoledge fruitful discussions with Jan Steinheimer, Yilun Du, Nan Su, Longgang Pang, Yingxun Zhang, and Horst Stoecker. The authors acknowledge
support by the computing server C3S2 in Huzhou University. The work is supported in part by the National Science Foundation of China Nos. U2032145, 11875125, and 12047568,
and the National Key Research and Development Program of China under Grant No. 2020YFE0202002, and the BMBF under the ErUM-Data project, and the AI grant at FIAS of SAMSON AG, Frankfurt, and the ``Ten Thousand Talent Program" of Zhejiang province (No. 2018R52017).




\begin{thebibliography}{0}

\bibitem{BALi08}
  B.~A.~Li, L.~W.~Chen, C.~M.~Ko,
  Phys.\ Rep.\  {\bf 464}, 113-281 (2008).

  \bibitem{Tsang:2012se}
  M. B. Tsang, J. R. Stone, F. Camera, P. Danielewicz, S. Gandolfi, K. Hebeler, C. J. Horowitz,
  Jenny Lee, W. G. Lynch, Z. Kohley, R. Lemmon, P. M\"{o}ller, T. Murakami, S. Riordan, X. Roca-Maza, F. Sammarruca, A. W. Steiner, I. Vida\~{n}a, and S. J. Yennello,
  Phys.\ Rev.\ C {\bf 86}, 015803 (2012).

\bibitem{Baldo:2016jhp}
  M.~Baldo and G.~F.~Burgio,
  Prog.\ Part.\ Nucl.\ Phys.\  {\bf 91}, 203 (2016).

\bibitem{Oertel:2016bki}
  M.~Oertel, M.~Hempel, T.~Kl\"{a}hn and S.~Typel,
  Rev.\ Mod.\ Phys.\  {\bf 89}, no. 1, 015007 (2017).

\bibitem{Li:2018lpy}
  B.~A.~Li, B.~J.~Cai, L.~W.~Chen and J.~Xu,
  Prog.\ Part.\ Nucl.\ Phys.\  {\bf 99}, 29 (2018).

\bibitem{Roca-Maza:2018ujj}
  X.~Roca-Maza and N.~Paar,
  Prog.\ Part.\ Nucl.\ Phys.\  {\bf 101}, 96 (2018).

\bibitem{Gao:2019vby}
  H.~Gao, S.~K.~Ai, Z.~J.~Cao, B.~Zhang, Z.~Y.~Zhu, A.~Li, N.~B.~Zhang and A.~Bauswein,
  Front.\ Phys.\ (Beijing) {\bf 15}, no. 2, 24603 (2020).


\bibitem{Ono:2019jxm}
  A.~Ono,
  Prog.\ Part.\ Nucl.\ Phys.\  {\bf 105}, 139 (2019).

\bibitem{Xu:2019hqg}
  J.~Xu,
  Prog.\ Part.\ Nucl.\ Phys.\  {\bf 106}, 312 (2019).
    Eur.\ Phys.\ J.\ A.\ {\bf 55}, 117 (2019).
\bibitem{Colonna:2020euy}
  M.~Colonna,
  Prog. Part. Nucl. Phys. {\bf 113}, 103775 (2020).
\bibitem{FOP-wyj}
 Y.~J.~Wang, Q.~F.~Li,
 Front.\ Phys.  {\bf 15}, 44302 (2020).
\bibitem{FOP-zyx}
 Y.~X.~Zhang, N.~Wang, Q.~F.~Li {\it et al.},
 Front.\ Phys. {\bf 15}, 54301 (2020).
\bibitem{Ma:2018wtw}
  C.~W.~Ma and Y.~G.~Ma,
  Prog.\ Part.\ Nucl.\ Phys.\  {\bf 99}, 120 (2018).

\bibitem{Feng:2018emx}
Z.~Q.~Feng,
Nucl. Sci. Tech. \textbf{29}, no.3, 40 (2018)
doi:10.1007/s41365-018-0379-z
[arXiv:1802.10294 [nucl-th]].



\bibitem{ROMP}
  C.~Giuseppe, C.~Ignacio, C.~Kyle {\it et al.},
  Rev.\ Mod.\ Phys.\ {\bf 91}, 045002 (2019).

\bibitem{XLF}
  X.~L.~Fang, J.~Li, X.~Li {\it et al.},
  Sci.\ China Phys.\ Mech.\ Astron. {\bf 62}, 969512 (2019).
\bibitem{HL}
  H.~Luo, M.~X.~Luo, K.~Wang {\it et al.},
  Sci.\ China Phys.\ Mech.\ Astron. {\bf 62}, 991011 (2019).
\bibitem{Nature1}
  Radovic.~A, Williams, Rousseau.~D {\it et al.},
  Nature {\bf 560}, 41 (2018).

\bibitem{Steinheimer:2019iso}
J.~Steinheimer, L.~Pang, K.~Zhou, V.~Koch, J.~Randrup and H.~Stoecker,
JHEP \textbf{12}, 122 (2019).
\bibitem{Song:2021rmm}
Y.~D.~Song, R.~Wang, Y.~G.~Ma, X.~G.~Deng and H.~L.~Liu,
Phys. Lett. B \textbf{814}, 136084 (2021)
doi:10.1016/j.physletb.2021.136084
[arXiv:2101.10613 [nucl-th]].
\bibitem{Wang:2020tgb}
R.~Wang, Y.~G.~Ma, R.~Wada, L.~W.~Chen, W.~B.~He, H.~L.~Liu and K.~J.~Sun,
Phys. Rev. Res. \textbf{2}, no.4, 043202 (2020)


\bibitem{Ma:2020bic}
C.~W.~Ma, D.~Peng, H.~L.~Wei, Y.~T.~Wang and J.~Pu,
Chin. Phys. C \textbf{44}, no.12, 124107 (2020).

\bibitem{Ma:2020mbd}
C.~W.~Ma, D.~Peng, H.~L.~Wei, Z.~M.~Niu, Y.~T.~Wang and R.~Wada,
Chin. Phys. C \textbf{44}, no.1, 014104 (2020).

\bibitem{PJC}
  Z.~A.~Wang, J.~C.~Pei, Y.~Liu,
  Phys.\ Rev.\ Lett {\bf 123}, 122501 (2019).
\bibitem{Utama:2016tcl}
R.~Utama, W.~C.~Chen and J.~Piekarewicz,
J. Phys. G \textbf{43}, no.11, 114002 (2016).

\bibitem{Zhou:2018ill}
  K.~Zhou, G.~Endr?di, L.~G.~Pang and H.~St?cker,
  Phys.\ Rev.\ D {\bf 100}, no. 1, 011501 (2019)
  doi:10.1103/PhysRevD.100.011501
\bibitem{Shi:2021qri}
  S.~Shi, K.~Zhou, J.~Zhao, S.~Mukherjee and P.~Zhuang,
  arXiv:2105.07862 [hep-ph].


\bibitem{CDPRC}
  C.~David, M.~Freslier, J.~Aichelin,
  Phys.\ Rev.\ C {\bf 51}, 1453 (1995).
\bibitem{SBASS}
  S.~A.~Bass, A.~Bischoff, J.~A.~Maruhn, H. St\"{o}cker, and W. Greiner,
   Phys.\ Rev.\ C {\bf 53}, 2358 (1996); S. A. Bass, A. Bischoff, C. Hartnack, J. A. Maruhn, J. Reinhardt, H. St\"{o}cker, and W. Greiner, J. Phys. G {\bf 20}, L21 (1994).
\bibitem{Haddad}
  F.~Haddad, K.~Hagel, J.~Li {\it et al.},
  Phys.\ Rev.\ C {\bf 55}, 1371 (1997).
\bibitem{JD}
  J.~D.~Sanctis, M.~Masotti, M.~Bruno {\it et al.},
  J.\ Phys.\ G {\bf 36}, 015101 (2008).

\bibitem{Li:2020qqn}
F.~Li, Y.~Wang, H.~L\"u, P.~Li, Q.~Li and F.~Liu,
J. Phys. G \textbf{47}, no.11, 115104 (2020).
\bibitem{lifup2}
F.~Li, Y.~Wang, Z.~Gao, P.~Li, H.~Lv, Q.~Li, C.~Y.~Tsang and M.~B.~Tsang,
[arXiv:2105.08912 [nucl-th]].
\bibitem{Kuttan:2020kha}
M.~Omana Kuttan, J.~Steinheimer, K.~Zhou, A.~Redelbach and H.~Stoecker,
Phys. Lett. B \textbf{811}, 135872 (2020).




\bibitem{PLG}
  L.~G.~Pang, K.~Zhou, N.~Su {\it et al.},
  Nat.\ Commun {\bf 9} 210 (2018).
\bibitem{Du:2019civ}
Y.~L.~Du, K.~Zhou, J.~Steinheimer, L.~G.~Pang, A.~Motornenko, H.~S.~Zong, X.~N.~Wang and H.~St\"ocker,
Eur. Phys. J. C \textbf{80}, no.6, 516 (2020)
doi:10.1140/epjc/s10052-020-8030-7
[arXiv:1910.11530 [hep-ph]].
\bibitem{Kvasiuk:2020izb}
Y.~Kvasiuk, E.~Zabrodin, L.~Bravina, I.~Didur and M.~Frolov,
JHEP \textbf{07}, 133 (2020).

\bibitem{Thaprasop:2020mzp}
  P.~Thaprasop, K.~Zhou, J.~Steinheimer and C.~Herold,
  Phys.\ Scripta {\bf 96}, no. 6, 064003 (2021)
  doi:10.1088/1402-4896/abf214

\bibitem{Zhao:2021yjo}
  Y.~S.~Zhao, L.~Wang, K.~Zhou and X.~G.~Huang,
  arXiv:2105.13761 [hep-ph].

\bibitem{SAB}
  S.~A.~Bass, M.~Belkacem, M.~Bleicher {\it et al.},
  Prog.\ Part.\ Nucl.\ Phys.\  {\bf 41}, 255 (1998).
\bibitem{BLE}
  M.~Bleicher, E.~Zabrodin, C.~Spieles {\it et al.},
  J.\ Phys.\ G {\bf 25}, 1859 (1999).
\bibitem{qfli1}
  Q.~F.~Li, C.~W.~Shen, C.~C~Guo {\it et al.},
  Phys.\ Rev.\ C {\bf 83}, 044617 (2011).
\bibitem{qfli2}
  Q.~F.~Li, G.~Graf, M.~Bleicher,
  Phys.\ Rev.\ C {\bf 85}, 034908 (2012).



\bibitem{Zhang:2006vb}
  Y.~Zhang and Z.~Li,
  Phys.\ Rev.\ C {\bf 74}, 014602 (2006).

  \bibitem{Wang:2013wca}
  Y.~Wang, C.~Guo, Q.~Li, H.~Zhang, Z.~Li and W.~Trautmann,
  Phys.\ Rev.\ C {\bf 89}, 034606 (2014).
\bibitem{Aichelin:1991xy} J. Aichelin, Phys. Rept.  {\bf 202}, 233 (1991).
  \bibitem{Hartnack:1997ez} C. Hartnack, R. K. Puri, J. Aichelin, J. Konopka, S. A. Bass, H. Stoecker and W. Greiner,  Eur. Phys. J.  A {\bf 1}, 151 (1998).

\bibitem{Wang:2020dru}
Y.~Wang, Q.~Li, Y.~Leifels and A.~Le F\`evre,
Phys. Lett. B \textbf{802}, 135249 (2020).

\bibitem{Wang:2014rva}
Y.~Wang, C.~Guo, Q.~Li, H.~Zhang, Y.~Leifels and W.~Trautmann,
Phys. Rev. C \textbf{89}, no.4, 044603 (2014).

\end{thebibliography}
\end{document}